\documentclass[preprint,aps,floats,amssymb,showpacs]{revtex4}
\usepackage{epsfig}

\newcommand{\be}{\begin{equation}}
\newcommand{\ee}{\end{equation}}
\newcommand{\bea}{\begin{eqnarray}}
\newcommand{\eea}{\end{eqnarray}}
\newcommand{\bean}{\begin{eqnarray*}}

\newcommand{\eean}{\end{eqnarray*}}
\font\upright=cmu10 scaled\magstep1 \font\sans=cmss10
\newcommand{\ssf}{\sans}
\newcommand{\stroke}{\vrule height8pt width0.4pt depth-0.1pt}
\newcommand{\Z}{\hbox{\upright\rlap{\ssf Z}\kern 2.7pt {\ssf Z}}}

\newcommand{\C}{{\rlap{\rlap{C}\kern 3.8pt\stroke}\phantom{C}}}
\newcommand{\R}{\hbox{\upright\rlap{I}\kern 1.7pt R}}
\newcommand{\CP}{\C{\upright\rlap{I}\kern 1.5pt P}}
\newcommand{\PP}{\hbox{\upright\rlap{I}\kern 1.5pt P}}

\newcommand{\identity}{{\upright\rlap{1}\kern 2.0pt 1}}

\newcommand{\pp}{\Delta}
\newcommand{\HH}{\mbox{\hbox{\upright\rlap{I}\kern 1.7pt H}}}

\newcommand{\zb}{{\bar z}}

\newcommand{\fr}{\frac}
\newcommand{\lm}{\lambda}

\newcommand{\al}{\alpha}
\newcommand{\bt}{\beta}
\newcommand{\pr}{\partial}
\newcommand{\hs}{\hspace{5mm}}

\newcommand{\dg}{\dagger}

\begin{document}




\title{Gravitating $SU(N)$ Monopoles from Harmonic
maps}
\renewcommand{\thefootnote}{\fnsymbol{footnote}}
\author{Yves Brihaye \footnote{yves.brihaye@umh.ac.be}}
\address{Facult\'e des Sciences, Universit\'e de Mons, 7000 Mons, Belgium}
\author{Betti Hartmann \footnote{b.hartmann@iu-bremen.de}}
\address{School of Engineering and Sciences, International
University Bremen (IUB), 28275 Bremen, Germany}
\author{Theodora Ioannidou \footnote{T.Ioannidou@ukc.ac.uk}}
\address{Institute of Mathematics,  University
of Kent, Canterbury CT2 7NF, UK   }
\author{Wojtek Zakrzewski\footnote{W.J.Zakrzewski@durham.ac.uk}}
\address{Department of Mathematical Sciences, University
of Durham, Durham DH1 3LE, U.K.}
\date{\today}
\setlength{\footnotesep}{0.5\footnotesep}

\begin{abstract}
Spherically symmetric solutions of the $SU(N)$ Einstein-Yang-Mills-Higgs system
are constructed  using the harmonic map ansatz.
This way the problem  reduces  to solving  a set of ordinary 
differential equations  for the appropriate profile functions. 
In the $SU(2)$ case, we recover
the equations studied in great detail previously, while in  the $SU(N)$   
($N  > 2$) one we find new solutions which correspond to monopole-antimonopole 
configurations.

\end{abstract}

\maketitle
\renewcommand{\thefootnote}{\arabic{footnote}}

\section{Introduction}
Topological defects \cite{vs} are thought to have formed during 
the phase transitions
that took place in the early universe. If a symmetry is spontaneously broken
($G\rightarrow H$),  the topology of the vacuum manifold ${\cal M}=G/H$
determines
which type of the defect arises during the symmmetry
breakdown. This way we could get domain walls, cosmic strings, monopoles
and textures if ${\cal M}$ is (respectively)  disconnected,
has contractible loops,
non-contractible 2-spheres and non-contractible 3-spheres.
In 1974, 't Hooft and Polyakov \cite{tp} realised that the bosonic part of the
 Georgi-Glashow
model, which is essentially an $SU(2)$ Yang-Mills-Higgs system, possesses
soliton solutions which, due to their topological properties, carry a magnetic
 charge.
Since the unbroken $U(1)$ group is associated with
the electromagnetic field, these solutions 
are said to be describing ``magnetic monopoles''. 
Soon afterwards people started looking at various embeddings of $SU(2)$
 into higher
gauge groups \cite{cofn,bn1,dt1}. The embedding into $SU(3)$ was first
studied in \cite{cofn} and the two possible embeddings which correspond
to a $SU(2)$, respectively $SO(3)$ subgroup, were investigated. 
For the latter embedding, solutions with magnetic charge $\pm\sqrt{3}$
 \cite{bw} as well as solutions with zero topological charge \cite{burz}
 were constructed. A systematic analysis of the solutions in an $SU(3)$
model with  a non-vanishing potential has been done in \cite{kunz}.
Recently, static monopole solutions of the second order $SU(N)$
BPS Yang-Mills-Higgs equations, which are not solutions of the first order
Bogomonlyi  equations, have also been constructed \cite{IS}.
These spherically symmetric solutions may be interpreted as 
monopole-antimonopole configurations and their construction involves 
the use of harmonic 
maps into complex projective spaces.

When an $SU(2)$ Yang-Mills-Higgs system is minimally coupled to gravity,
three different types of solutions are possible; namely embedded 
Reissner-Nordstr\"om solutions \cite{br}, gravitating monopoles
and non-abelian black holes \cite{bfm}. Gravitating monopoles exist only
up to a maximal value of the gravitational coupling beyond which their
Schwarzschild radius becomes larger than the radius of the monopole core.
At this maximal value of the gravitational coupling the solutions bifurcate
producing a branch of extremal Reissner-Nordstr\"om solutions.
Non-abelian black holes can be thought of as black holes situated inside the core
of a magnetic monopole. Consequently, they exist only in a limited domain
of the gravitational coupling-horizon plane. 
Note that gravitating monopoles in $SU(3)$ corresponding to a $SU(2)$ subgroup
have been studied in \cite{bp}, while the gravitating monopoles and non-abelian
black holes in $SU(5)$, corresponding to a $SU(2)$ subgroup, were studied in 
\cite{bh}.
Here we will construct the spherical symmetric 
gravitating solutions of the second order $SU(N)$ Einstein-Yang-Mills-Higgs
equations which are neither solutions of the first order Bogomolnyi equations nor 
simple embedding of the $SU(2)$ ones.

When one considers fields in three dimensional space, it is sometimes convenient
to introduce spherical polar coordinates to describe all points in space
with the origin of coordinate system located at a specific point - like
the centre of the soliton (e.g. monopole). Then one introduces
a radial variable $r$ and two angular variables describing points on the 
sphere of radius $r$ - the complex
variables $z$ and $\bar z$ (discussed later in more detail). 
Therefore, for a given $r$, we have maps from $S\sp2$ of radius $r$,
which we can interpret as compactified $R\sp2$. The 
harmonic map ansatz \cite{manton}
exploits this property; it uses maps of $S\sp2 \rightarrow G$ to construct
maps of $R\sp3 \rightarrow G$. 
This is done by considering maps
of $R\sp2\rightarrow G$ and assuming that the parameters of these maps
are functions of $r$ only. 
This cannot be done in an arbitrary way; the fact that
the resultant maps are those of $R\sp3 \rightarrow G$ imposes some
constraints of continuity etc (discussed in the next section).

Our paper is organized as follows. In Section II we present the model and 
introduce the harmonic map ansatz. 
In Section III we derive and then discuss the
 resulting equations for the $SU(2)$ case
while in Section IV the equations for the $SU(3)$ model 
 and also present
our numerical results. Finally,  in Section V we presents  our conclusions.

\section{The model}
The $SU(N)$ Einstein-Yang-Mills-Higgs action is given by:
 \be
  S=\int \left[\fr{R}{16\pi G}-\fr{1}{2} \mbox{tr}\left(F_{\mu
  \nu}\,F^{\mu  \nu}\right)-\fr{1}{4}\mbox{tr}\left(D_\mu \Phi\, D^\mu
  \Phi\right)+\fr{1}{8} \lm\,
  \left(\mbox{tr}\left(\Phi^2-\eta^2\right)\right)^2\right]\sqrt{-g}\, d^4x
  \label{ac}
  \ee
  where $g$ denotes the determinant of the
metric while the field strength tensor is defined by: 
\be
F_{\mu
  \nu}=\pr_\mu A_\nu-\pr_\nu A_\mu+[A_\mu,A_\nu]
\ee
and the
  covariant derivative of the Higgs field reads: 
\be
D_\mu \Phi=\pr_\mu
  \Phi+[A_\mu,\Phi].
\ee 

The matrix $\eta$ represents a constant matrix of the form:
$\eta=i v {\bf 1}_N$, where $v \in {\mathbb R}$ and  ${\bf 1}_N$ denotes the
unit matrix in $N$ dimensions. 
  The constants in the action represent Newton's constant $G$,
 the Higgs self-coupling
  constant $\lm$ and the vacuum expectation value of the Higgs
  field $v$. Note that for $N >2$ the potential in (\ref{ac}) is not the
most general one that could have been used.

Variation of the action (\ref{ac}) with respect to the metric
$g^{\mu \nu}$ leads to the Einstein equations
 \be
 R_{\mu \nu}-\fr{1}{2}g_{\mu \nu} R=8\pi G \,T_{\mu \nu}
 \label{En}
 \ee
 with the stress-energy tensor   $T_{\mu\nu}=g_{\mu\nu}{\cal L}-2\fr{\pr{\cal L}}
{\pr g^{\mu\nu}}$ given by
 \bea
 T_{\mu \nu}&=&\fr{1}{2}\mbox{tr}\left(D_\mu \Phi\, D_\nu \Phi-
 \fr{1}{2}g_{\mu \nu} D_\al\Phi\,D^\al \Phi\right)
 +2\mbox{tr}\left(g^{\al \bt}\,F_{\mu \al}F_{\nu \bt}-\fr{1}{4}
g_{\mu \nu}\,F_{\al \bt}F^{\al \bt}\right)\nonumber\\
&&+\fr{1}{8}g_{\mu \nu} \lm
\,\left(\mbox{tr}\left(\Phi^2-\eta^2\right)\right)^2.
 \eea

Variation with respect to the gauge fields $A_\mu$ and the Higgs
field $\Phi$ leads to the matter equations
 \bea
\fr{1}{\sqrt{-g}}\,D_\mu\left(\sqrt{-g}\,F^{\mu\nu}\right)-\fr{1}{4}
\left[\Phi,D^\nu \Phi\right]&=&0, \nonumber\\
\fr{1}{\sqrt{-g}}\,D_\mu\left(\sqrt{-g}\,D^\mu
\Phi\right)+\lambda\,\mbox{tr}\left(\Phi^2-\eta^2\right)\,
 \Phi&=&0   \ . \label{eq}
 \eea

 In what follows we consider the static Einstein-Yang-Mills-Higgs equations 
in order to construct  their spherically symmetric and purely magnetic
 (ie $A_0=0$)  solutions based on the harmonic map ansatz first introduced in
\cite{IS}.

\subsection{The Harmonic Map Ansatz}

The starting point of our investigation is the introduction of the
coordinates $r,z,\zb$ on $\R^3$. In terms of the usual spherical
coordinates $r,\theta,\phi$ the Riemann sphere variable $z$ is given by
$z=e^{i\phi} \tan(\theta/2)$ and $\bar{z}$ is the complex
conjugate of $z$. In this system of coordinates the
Schwarzschild-like metric reads:
\begin{equation}
\label{metric}
ds^2=-A^2(r)B(r)dt^2+\fr{1}{B(r)}dr^2+\frac{4r^2}{(1+|z|^2)^2}
\,dz d\bar{z}, \hs B(r)=1-\frac{2m(r)}{r},
 \label{s}
\end{equation}
where  $A$ and $B$ are the metric  functions which are real 
and depend only on the radial coordinate $r$, 
and
$m(r)$ is the mass function.
 The (dimensionfull) mass of the solution
is given by $m_{\infty}\equiv m(\infty)$. 
For this metric the square-root of the determinant takes the simple form:
\begin{equation}
\sqrt{-g}=iA(r)\,\frac{2r^2}{(1+|z|^2)^2}.
\end{equation}
 Using (\ref{s}) the matter equations (\ref{eq}) read:
 \bea
&&\hspace{-10mm}\fr{1}{4}\left[D_r \Phi,\Phi\right]
-\fr{(1+|z|^2)^2}{2r^2} \left(D_z F_{r\zb}+D_{\zb}
F_{rz}\right)=0, \label{en}\\
&&\hspace{-10mm}\fr{1}{4}\,\left[D_z \Phi,\Phi\right]+
\fr{1}{A}D_r\left(A B\,F_{rz}\right)-\fr{1}{2r^2}\,
D_z\left((1+|z|^2)^2\,F_{z\zb} \right)=0,
\label{d}\\
&&\hspace{-10mm}\fr{1}{A r^2} D_r \left(A B\,r^2
D_r\Phi\right)+\fr{ (1+|z|^2)^2}{ 2r^2}\left(D_z
D_{\zb} \Phi+D_{\zb}D_z \Phi
\right)+\lambda\,\mbox{tr}\left(\Phi^2-\eta^2\right)\, \Phi=0\hs\hs\label{t}
 \eea
 In addition, the Einstein equations (\ref{En}) take the form:
 \begin{equation}
 \frac{2}{r^2}m' = 8\pi
 G\, T^0_0,\hs  \hs
 \fr{2}{r} \fr{A'}{A}\,B =
8\pi G \,\left( T^0_0-T^r_r \right)
\label{ff}
 \end{equation}
where the prime denotes the derivative with respect to $r$ and
\bea
T^0_0&=&-
 \fr{B(1+|z|^2)^2}{r^2}\mbox{tr}\left(|F_{rz}|^2\right)+
\fr{(1+|z|^2)^4}{4r^4}\mbox{tr}\left(F_{z\bar{z}}^2\right) -
\frac{B}{4} \mbox{tr}\left( (D_r \Phi)^2\right)\nonumber \\
&-&\fr{(1+|z|^2)^2}{4r^2}\mbox{tr}\left(|D_z\Phi|^2\right)
+\fr{\lm}{8}\,\left(\mbox{tr}\left(\Phi^2
-\eta^2\right)\right)^2  \ , \
\eea
\bea
T^0_0-T^r_r& =&
-\frac{2B(1+|z|^2)^2}{r^2}\mbox{tr}\left(|F_{rz}|^2\right) 
- \frac{B}{2} \mbox{tr}
\left((D_r\Phi)^2\right).
 \eea

Next we introduce the following ansatz \cite{IS} for
 the $SU(N)$ gravitating monopoles:
 \be
 \Phi=i\sum_{j=0}^{N-2} h_j\left(P_j-\fr{1}{N}\right),\hs \hs
 A_z=\sum_{j=0}^{N-2}g_j\left[P_j,\pr_z P_j\right],\hs
 A_r=0 \label{f}
 \ee
 where $h_j(r)$, $g_j(r)$ are the radial dependended matter profile functions 
and $P(z,\zb)$
 are $N\times N$ Hermitian  projectors: $P_j=P_j^\dg=P_j^2$,
 which are independent of the radius $r$.
 Note that all $N-1$ projectors $P_i$ are  orthogonal to each other since
 $P_iP_j=0$ for $i \neq j$ and 
that we are working in a real gauge, since
 $A_{\zb}=-A_z^\dg$.

As shown in \cite{IS}, the projectors $P_k$ defined as 
\be P_k=\fr{(\pp^k f)^\dg \pp^k f}{|\pp^k f|^2},
 \hs \hs k=0,..,N-1 
\ee
where 
\be
  \pp f=\pr_z f- \fr{f \,(f^\dg\,\pr_z f)}{|f|^2}
 \ee
give us our required set of orthogonal harmonic maps (for
details see \cite{Za}).
Moreover, the   harmonic maps with spherical symmetry
can obtained by applying the
 orthogonalization procedure to the
initial holomorphic vector  $f$ given by 
\be f=(f_0,...,f_j,...,f_{N-1})^t, \ \
\mbox{where} \ \ f_j=z^j\sqrt{{N-1}\choose j} \label{smap} \ee and
${N-1}\choose j$ denote the binomial coefficients. 
Then equation (\ref{en}) is automatically satisfied.

In dealing with the equations which arise from the harmonic map
ansatz (\ref{f}) it is convenient to replace the profile
functions $h_j(r)$, $g_j(r)$ by the functions $b_j(r)$, $c_j(r)$ which
are defined as the following linear combinations of $g_j$ and $b_k$:
 \be
  h_j=\sum_{k=j}^{N-2}b_k,\hs c_j=1-g_j-g_{j+1}, \hs j=0,\dots,N-2,
 \ee
where $g_{N-1}=0$.

Next we will describe in  detail the gravitating
monopoles obtained from our harmonic map ansatz  for
the simplest cases of $SU(2)$ and $SU(3)$. The situation for
general $SU(N)$ will then become clear.

\section{ Gravitating monopoles in $SU(2)$}

For $N=2$ there are two profile functions, $b_0,c_0$ and
our ansatz (\ref{f}) reduces equations (\ref{d}) and (\ref{t}) to
the following set of second order nonlinear ordinary differential
equations:
 \bea
\fr{1}{A}\left(AB\,c_0'\right)'&=&
\fr{1}{4}\,b_0^2\,c_0+\fr{1}{r^2}\,c_0\left(c_0^2-1\right), \\
\fr{1}{Ar^2}\left(r^2\,AB\,b_0'\right)'
&=& \fr{2}{r^2}\,b_0c_0^2+\fr{\lm}{2}\,b_0
\left(b_0^2-4v^2\right).
 \eea
Finally the Einstein equations (\ref{ff}) take the form:
\bea
\fr{2}{r^2}m'&=&8\pi G\,\left[\fr{B}{8}\, b_0'^2+\fr{1}{4r^2}\,b_0^2\,c_0^2
+\fr{1}{r^2}\,Bc_0'^2+\fr{1}{2r^4}\left(1-c_0^2\right)^2 \right. 
+ \left. \fr{\lm}{16}
\left(b_0^2-4v^2\right)^2\right]  \ , \\
\fr{2}{r}\fr{A'}{A}&=&8\pi G\,\left(\fr{1}{4}\,b_0'^2+\fr{2}{r^2}\,
c_0'^2\right) \ ,
\eea
where $m(r)$ is given by (\ref{metric}).
These equations have previously been studied in great detail in
\cite{bfm} (after the rescale of the Higgs profile function
$b_0 \rightarrow 2b_0$). We will not repeat the numerical calculations here,
 but refer
the reader to the mentioned papers.

\section{Gravitating monopoles in $SU(3)$}

For $N=3$ there are four profile functions, $b_0,b_1,c_0,c_1$ and
our ansatz (\ref{f}) reduces equations (\ref{d}) and (\ref{t}) to
the following set of second order nonlinear ordinary differential
equations:
 \bea
\fr{1}{A}\left(AB\,c_j'\right)'&=&\fr{1}{4}\,b_j^2c_j-\fr{1}{r^2}c_j\left
(1-2c_j^2+c_k^2\right)\label{beq} \ , \\
\fr{1}{Ar^2}\left(r^2\,ABb_j'\right)'
&=&\fr{2}{r^2}
\left(2b_jc_j^2-b_kc_k^2\right)+\fr{2\lm}{3}\,b_j
\left(b_j^2+b_kb_j+b_k^2-\fr{9}{2}v^2\right)\hs\hs\hs \label{ceq}   \ .
 \eea
Here the indices are chosen from the set $\{0,1\}$, $k\neq j$,
and we assume the symmetry under the interchange of indices
$0\leftrightarrow 1$ when applied to both the $b_j$ and $c_j$
functions. 
Note that in the flat limit, ie for $A=B=1$, and for
$\lambda=0$ the equations of \cite{IS} are recovered.

Finally, the Einstein equations (\ref{ff}) take the form:
\bea
\fr{2}{r^2}m'&=&8\pi G\,\bigg[\fr{B}{6} \left(b_0'^2+b_0'b_1'+b_1'^2\right)
+\fr{1}{2r^2}\left(b_0^2c_0^2+b_1^2c_1^2\right)
+\fr{2B}{r^2}\left(c_0'^2+c_1'^2\right)\nonumber\\
&+&\fr{2}{r^4}\left(1-c_0^2-c_1^2+c_0^4+c_1^4-c_0^2c_1^2\right)
+\fr{\lm}{18}
\left(b_0^2+b_0b_1+b_1^2-\fr{9}{2}v^2
\right)^2\bigg],
\label{meq}  \\
\fr{2}{r}\fr{A'}{A}&=&8\pi G\,\left[\fr{1}{3}\left(b_0'^2+b_0'b_1'+b_1'^2
\right)+\fr{4}{r^2}\left(c_0'^2+c_1'^2\right)\right]. \label{aeq}
\eea
These equations correspond to the $(\tau_1,\tau_2,\tau_3)=
(\lambda_7,-\lambda_5,\lambda_2)$ embedding of $SU(2)$ into $SU(3)$ (see section
below).
The $\tau_i$'s here are the $SU(2)$ (or $SO(3)$) generators, while the 
$\lambda_i$'s denote
the Gell-Mann matrices. 

The equations corresponding to the $(\tau_1,\tau_2,\tau_3)=
(\lambda_1/2,\lambda_2/2,\lambda_3/2)$ embedding of
$SU(2)$ into $SU(3)$ have been studied in \cite{bp}
and are not included in our approach as can be seen  by setting 
either $c_0$ or $c_1$  equal to zero and comparing
our equations with those of \cite{bp}.
This is due to the fact that in our construction the corresponding 
Higgs and gauge fields are non-embeddings of the $SU(2)$ ones.

\subsection{Comparison with the spherically symmetric ansatz}

The spherically symmetric ansatz used in \cite{burz,kunz,br} is
 given as follows:
\begin{equation}
\Phi= F_1(r) {\bf Y} + F_2(r)\left({\bf Y}^2-\frac{2}{3}\right ),
\end{equation}
where ${\bf Y}=\hat{r}\cdot \vec{\Lambda}$, $\vec{\Lambda}=(\lambda_7,
-\lambda_5,
\lambda_2)$ (or in components $Y_{ab}=-i\epsilon_{abc} \hat{r}_c$) and 
\begin{equation}
A_i=\frac{G(r)}{2r}(\vec{r} \times \Lambda)_i 
+ \frac{H(r)}{2r}\left[(\hat{r}\times\vec{\Lambda})_i,
\hat{r}\cdot \vec{\Lambda}\right]_+ , \hs \hs A_0=0, 
\end{equation}
where $[$ $,$ $]_+$ denotes the anti-commutator.

After some algebra, it can be seen that the above Higgs field can be rewritten
in terms of hermitian and orthogonal projectors constructed out of the 
matrix ${\bf Y}$
 (note that ${\bf Y}^3={\bf Y}$):
\begin{equation}
\Phi=(b_0+b_1)\left(P_0-\frac{1}{3}\right)+b_1\left(P_1-\frac{1}{3}\right),
\end{equation}
with 
\begin{equation}
P_0=\frac{{\bf Y}+{\bf Y}^2}{2}, \hs \hs P_1={\bf Y}^2 -1.
\end{equation}
With these projectors, the gauge fields can be constructed in an analogous way.
Finally, the full correspondence between the harmonic map ansatz and the
spherically symmetric ansatz can be established by observing that
\begin{equation}
F_1=\frac{1}{2}(b_0+b_1), \hs F_2=\frac{1}{2}(b_0-b_1), \hs
\tilde{G}\equiv G-2=c_0-c_1, \hs H=c_0+c_1.
\end{equation}

\subsection{Numerical results}

As in  \cite{bfm}, without any loss of generality,
we set the vacuum expectation value $v=1$ (different
$v$'s can always be accommodated by rescaling the radial coordinate
and the Higgs field functions) and we  also define
$\alpha^2\equiv 4\pi G$. Then the dimensionless mass of the solutions
is defined as: $M=\frac{m_{\infty}}{\alpha^2}$.
Here  we will consider only the $\lambda=0$  case and
 leave $\lambda\ne 0$ to a further study.	  

In this section we  construct  the monopole solutions
of equations (\ref{beq})-(\ref{aeq}) numerically using a collocation method
for the boundary-value ordinary differential equations \cite{acr}. In this procedure
the set of non-linear coupled differential equations is solved using a damped
Newton method of quasi-linearization.

 The boundary conditions for the
metric profiles read:
\begin{equation}
m(r=0)=0 \ ,\hs \hs \hs \ A(r=\infty)=1 \ 
\end{equation}
and we also assume that $A(r=0)$ is finite.

Since the equations (\ref{beq})-(\ref{aeq}) are very similar to those
 of the $SU(2)$ case, we expect that their solutions will bifurcate
into extremal Reissner-Nordstr\"om solutions at some critical value
of the gravitational coupling constant $\alpha$. Prior to describing 
our numerical
results, it is worth mentioning that the charge of the limiting 
Reissner-Nordstr\"om
solution can be given in terms of the asymptotic values of the gauge fields
$\tilde{c}_0\equiv c_0(\infty)$, $\tilde{c}_1\equiv c_1(\infty)$. 
Due to (\ref{meq}) and taking into account the asymptotic behaviour of the various fields,
we find that the mass function $m_{RN}(r)$ of the Reissner-Nordstr\"om 
solution is given by:
\begin{equation}
\label{mRN}
m_{RN}(r)=m_{\infty,RN}-\frac{\alpha^2 Q^2}{2r},  \hs \hs \hs
Q^2=4\left(1-\tilde{c}^2_0-\tilde{c}^2_1+\tilde{c}^4_0+\tilde{c}^4_1-
\tilde{c}^2_0\tilde{c}^2_1\right).
\end{equation}
Note that for the extremal Reissner-Nordstr\"om
 solution $m_{\infty,RN}=\alpha Q$.

Following the discussions in \cite{IS} we note that there are three 
different types of solutions which 
seem to be of particular interest. They can be distinguished from each
other by whether they satisfy, or not, the Bogomolnyi equation and
whether the symmetry-breaking 
(SB) is maximal (unbroken group $U(1)\times U(1)$) or minimal
(unbroken group $U(2)$). 

First, we will discuss in detail
the case of non-Bogomolnyi maximal SB solutions which satisfy the 
following boundary
conditions for the fields:
\begin{equation}
b_0(r=0)= 0, \hs b_1(r=0)=0, \hs c_0(r=0)=1, \hs c_1(r=0)=1 
\end{equation}
at the origin  and 
\begin{equation}
b_0(r=\infty)=-2, \hs \hs b_1(r=\infty)=4, \hs\hs c_0(r=\infty)=0, 
\hs\hs c_1(r=\infty)=0 
\end{equation}
at infinity. Any of such solutions has a magnetic charge 
$(0,2)$ \cite{IS} and according
to the discussion given there can be interpreted as a superposition
of two monopoles and two pairs of monopoles-antimonopoles.
In the flat limit, this solution has a mass $M=4.5$.
When gravity is minimally coupled to the system, the solution gets
smoothly deformed by it. The metric function $B(r)$ develops a minimum
which gets deeper as  $\alpha$ increases. At a critical
value of $\alpha$ the solution develops a double zero at some finite
value of $r=r_h$, which can be interpreted as the horizon of the
extremal Reissner-Nordstr\"om  solution. Thus, the limiting solution
can be described by this field for $r \geq r_h$, while it
is non-trivial and non-singular for $0 \leq r < r_h$. 
This is illustrated in Fig.~1 (respectively 2), where we present the
profiles of the metric and gauge functions (respectively, metric and Higgs 
functions)
for the flat limit $\alpha=0.0$ and $\alpha\approx \alpha_{cr}=0.615$. 

As in the $SU(2)$
case, the numerical results indicate that the gravitating solution exists up    
to a maximal value of $\alpha$, $\alpha_{max}\approx 0.625$. By decreasing
$\alpha$ from $\alpha_{max}$, another branch of solutions can be constructed
 and  $B_{min}=0$ is reached at the critical value 
 $\alpha_{cr}\approx 0.615$. This is illustrated in Fig.~3, where plots of 
 $\frac{m_{\infty}}{\alpha}$ and of the minimal value $B_{min}$ in terms 
 of $\alpha$ are presented. Note that  at $\alpha=\alpha_{cr}$, the quantity 
$\frac{m_{\infty}}{\alpha}$ equals two.
 Since for the Reissner-Nordstr\"om  solution
$\frac{m_{\infty,RN}}{\alpha}=Q$, this implies that the solution
indeed bifurcates into a charge-two Reissner-Nordstr\"om
 solution (as can be calculated 
from (\ref{mRN})).

Next we studied the gravitating analogue of the Burzlaff  solution 
\cite{burz}. In this case $b_0=-b_1$ while $c_0=c_1$ and
 the solution corresponds
to a non-Bogomolnyi  non-maximal SB case with charge $(0,[2])$. 
[Here the notation is that magnetic weights are defined by square brackets 
\cite{IS}.]
We choose the following boundary conditions for the fields:
\begin{equation}
b_0(r=0)=0, \hs\hs b_1(r=0)=0, \hs\hs  c_0(r=0)=1, \hs\hs c_1(r=0)=1
\end{equation}
at the origin  and
\begin{equation}
b_0(r=\infty)=\sqrt{3}, \hs\hs b_1(r=\infty)=-\sqrt{3},\hs\hs 
 c_0(r=\infty)=0,\hs\hs c_1(r=\infty)=0
\end{equation}
at infinity. The choice of the boundary
conditions for the Higgs field at infinity is not fixed for a vanishing potential;
however,  following \cite{kunz}
we have chosen our conditions  as if the potential was present.
 This solution, due to (\ref{mRN}), should bifurcate
producing a charge-two Reissner-Nordstr\"om solution as confirmed
 in Fig. 3.
Again, we find a back-bending since the gravitating Burzlaff solution
exists up to $\alpha_{max}\approx 1.106$ and reaches $B_{min}=0$ at 
$\alpha=\alpha_{cr}\approx 1.063$. 
In the flat limit, the mass of the solution 
is $M=2.5$. Since the solutions are less heavy than
the ones in the non-Bogomolnyi maximal SB case, it is clear that
the critical value of the gravitational coupling should be bigger 
in the Burzlaff case (which  is confirmed by our numerical results).

Finally, we constructed the gravitating analogues of the 
Bais solution \cite{bw} by choosing the boundary conditions (again)
as if the potential were present \cite{kunz}.
The boundary conditions were
\begin{equation}
b_0(r=0)=0, \hs\hs b_1(r=0)=0, \hs\hs c_0(r=0)=1, \hs\hs c_1(r=0)=1
\end{equation}
at the origin  and
\begin{equation}
b_0(r=\infty)=-\sqrt{3}, \hs\hs b_1(r=\infty)=0, \hs\hs 
 c_0(r=\infty)=0, \hs\hs c_1(r=\infty)=\frac{1}{\sqrt{2}}
\end{equation}
at infinity. These solutions correspond 
to the Bogomolnyi non-maximal SB solutions
with charge $(2,[1])$. 
Due to (\ref{mRN}) the branch of solutions should bifurcate
producing a branch of Reissner-Nordstr\"om solutions with charge $\sqrt{3}$ 
as shown in Fig. 3.
 In fact  the gravitating Bais solutions exist up to
a maximal value of $\al$: $\alpha_{max}\approx 1.446$. We were not able to find
a back-bending and it is likely that there is none 
since $B_{min}\approx 4.7 \cdot 10^{-3}$ at $\alpha=\alpha_{max}$.
If there were  a back-bending, then the branch of solutions would be very small.
Since, in  the flat limit, the mass of the gravitating Bais solution (again)
 is smaller
than in the other two cases studied, the maximal value of $\alpha$ for
 the former is larger than for the latter.

\section{Conclusions}

We have studied gravitating $SU(N)$ monopoles by relying on the harmonic
 map ansatz.
In the $SU(2)$ case, we have recovered the equations studied previously 
in great detail
in \cite{bfm}; while in  the $SU(3)$ one, we have found the gravitating
 analogues of the
solutions obtained by the embedding of $SO(3)$ into $SU(3)$.
 Since in the case of the 
vanishing potential (considered here),
the boundary conditions of the Higgs profile functions 
 at infinity are not fixed, they can be chosen at will. 
In fact they have been chosen so that
 the
gravitating analogues of the solutions discussed in \cite{IS} can be  
constructed. These solutions correspond to  monopole-antimonopole
configurations and not to single monopoles as constructed in  \cite{bp}.

In all three cases studied here,  it has been found that the solutions bifurcate 
producing a branch
of extremal Reissner-Nordstr\"om ones with charge $Q$,  which is 
fixed by the asymptotic values of the gauge fields.
Interestingly, our numerical results indicate that a second branch of solutions,
which extends backwards from the maximal possible value of the gravitational
coupling, exists only in the case of non-maximal symmetry breaking. 
In the case of maximal symmetry breaking 
 (even in the limit of vanishing Higgs coupling) no second branch was found.
This is in contrast with the case of single  $SU(2)$ \cite{bfm}
and $SU(3)$ \cite{bp} monopoles where the second branch of solutions exists 
for
vanishing or small Higgs coupling. 

In this paper we have not constructed solutions for non-vanishing potential
or non-abelian black hole solutions. A systematic
study of these solutions is left to a future work \cite{bhiz}.
In particular, we would like to  study
the corresponding configurations for intermediate Higgs coupling constants.
It is known that for the single  $SU(2)$ and $SU(3)$  monopoles
 the so-called ``Lue-Weinberg'' \cite{lw} phenomenon was observed; ie
for intermediate values of the
Higgs coupling  the solutio ns develop a second ``inner'' horizon
and in the limit of critical gravitational coupling describe ``hairy'' 
black holes.
It will be interesting to see whether this phenomenon persists for 
monopole-antimonopole
configurations.

{\bf Acknowledgements} YB acknowledges the Belgian F.N.R.S. for financial
support. TI thanks K Kokkotas for useful discussions.

 \newpage
\begin{figure}
\centering
\epsfysize=18cm
\mbox{\epsffile{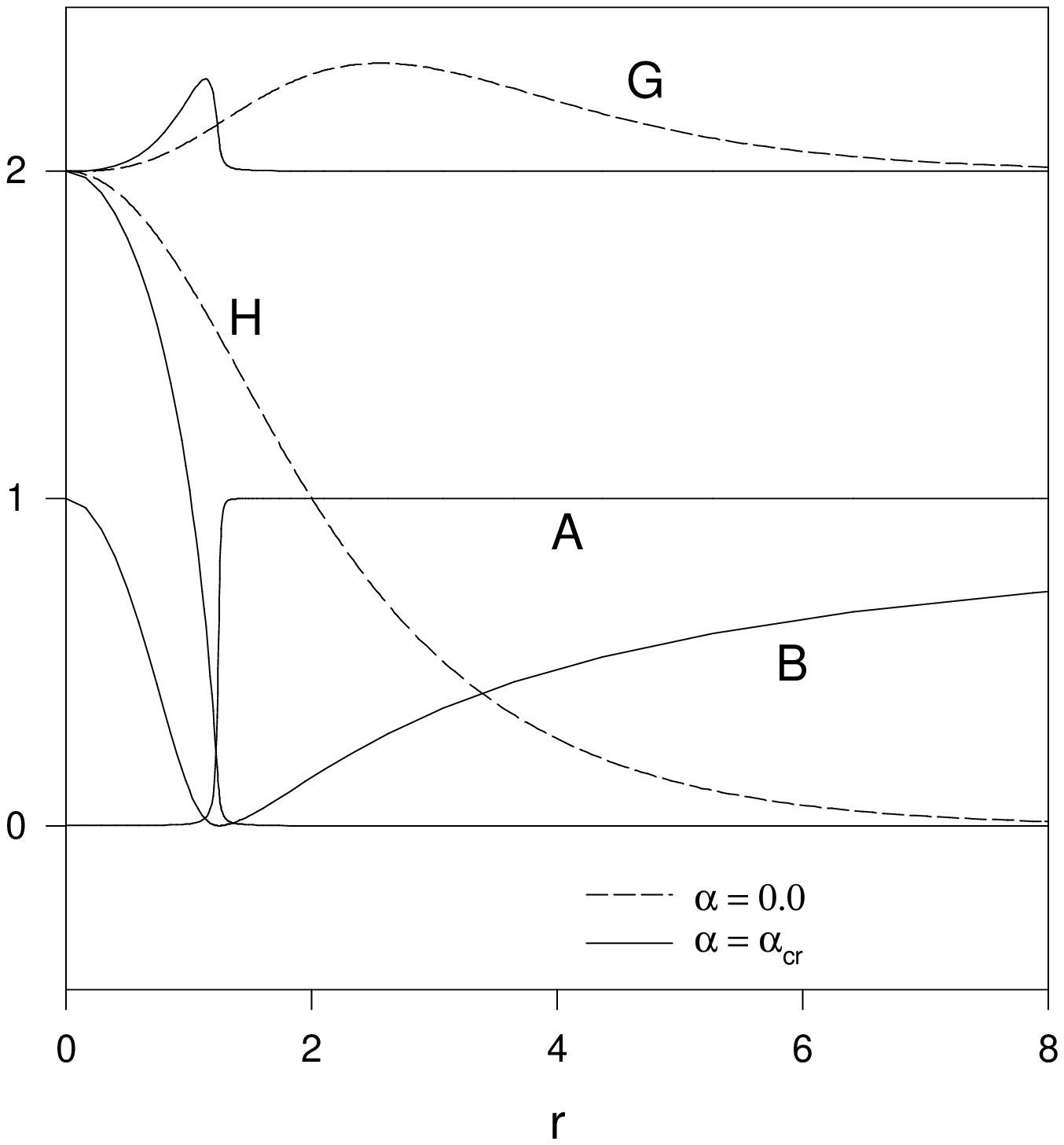}}
\caption{\label{Fig.1} The profiles of the metric  $A(r)$ and $B(r)$
as well as of the gauge fields  $G(r)=c_0(r)-c_1(r)+2$ and
$H(r)=c_0(r)+c_1(r)$ are presented for the  non-Bogomolnyi  maximal SB 
solutions
 in the flat limit $\alpha=0.0$ (dashed) and close to the critical limit 
$\alpha=\alpha_{cr}$ (solid).
Note that for $\alpha=0.0$,
$A(r)$ and $B(r)$ are constant since $A(r)=B(r)=1$.  }
\end{figure}
 \newpage
\begin{figure}
\centering
\epsfysize=18cm
\mbox{\epsffile{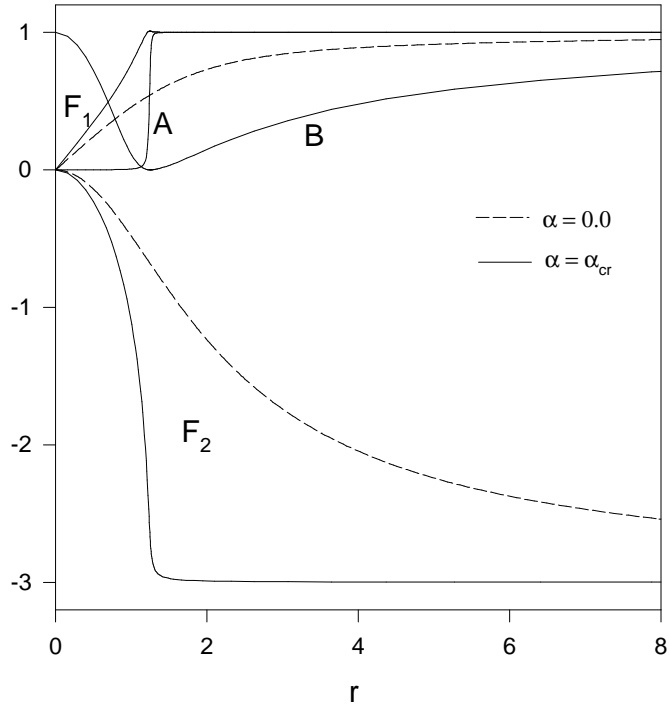}}
\caption{\label{Fig.2} Same as Fig. 1 for the profile functions of the 
 Higgs field 
 $F_1(r)=\frac{1}{2}(b_0(r)+b_1(r))$
and $F_2(r)=\frac{1}{2}(b_0(r)-b_1(r))$. For comparison, we show again
the metric functions $A(r)$ and $B(r)$.}
\end{figure}
 \newpage
\begin{figure}
\centering
\epsfysize=18cm
\mbox{\epsffile{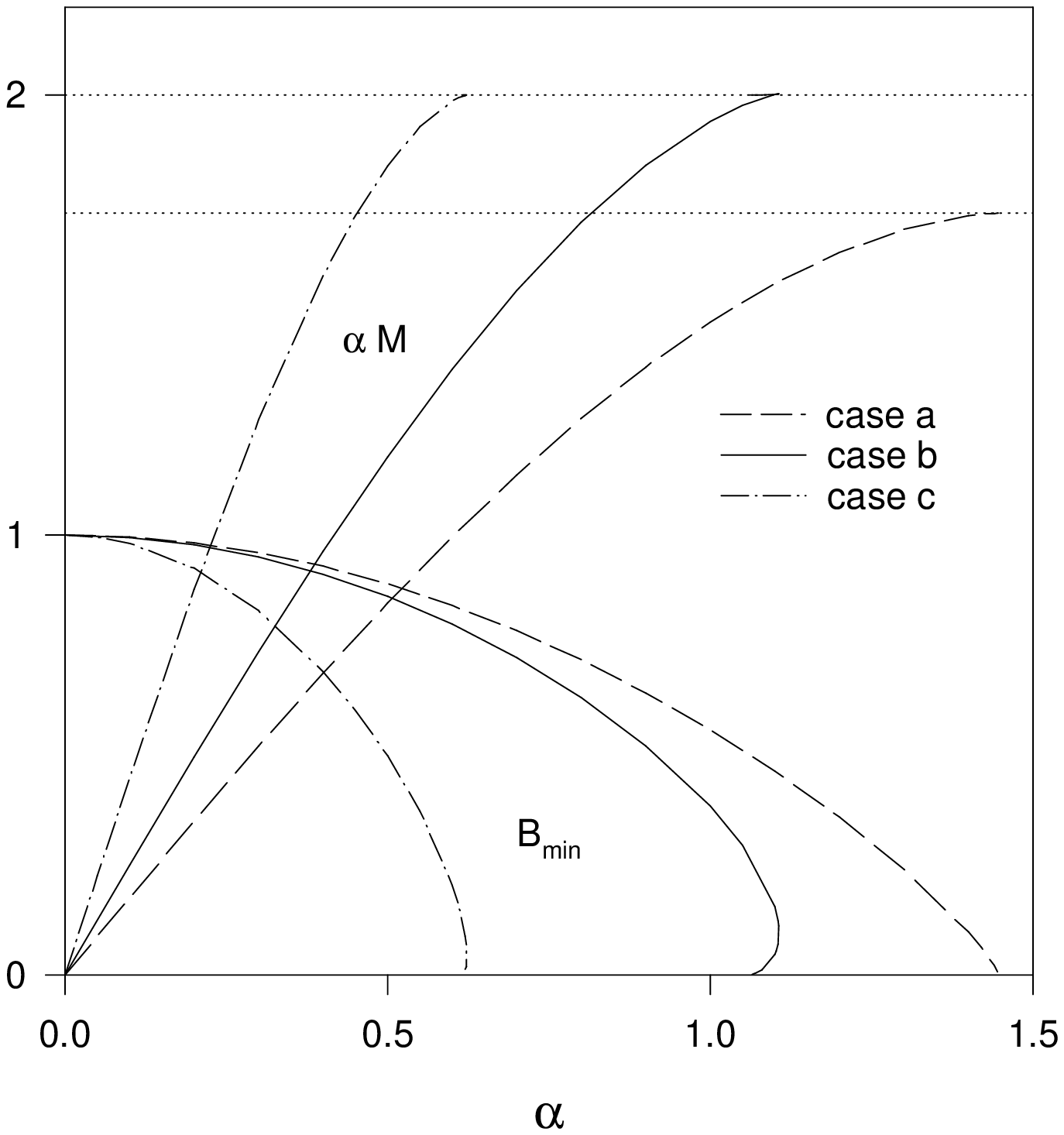}}
\caption{\label{Fig.3} The quantity $\alpha M=\frac{m_{\infty}}{\alpha}$ 
and 
the minimal value of the metric function $B(r)$, $B_{min}$, 
in terms of $\alpha$ is plotted
for a) the Bogomolnyi non-maximal SB  case 
(gravitating Bais solution), b) the non-Bogomolnyi
non-maximal SB case (gravitating Burzlaff solution) and 
c) the non-Bogomolnyi maximal SB case.}
\end{figure}

\begin{thebibliography}{99}
\bibitem{vs} A. Vilenkin and E. P. S. Shellard, {\it Cosmic strings and other topological
defects}, Cambridge University Press, 1994.
\bibitem{tp} G. 't Hooft, Nucl. Phys. {\bf B79} (1974), 276;
A. M. Polyakov, JETP Lett. {\bf 20} (1974), 194. 
\bibitem{cofn} E. Corrigan, D. I. Olive, D. Fairlie and J. Nuyts, 
Nucl. Phys. {\bf B 106} (1976), 475.
\bibitem{bn1} 
Y. Brihaye and J. Nuyts, 
J. Math. Phys. {\bf 18} (1977), 2177. 
\bibitem{dt1} 
C. P. Dokos and T. N. Tomaras, Phys. Rev. {\bf D 21} (1980), 2940.  
\bibitem{bw} F. A. Bais and H. A. Weldon, Phys. Rev. Lett. {\bf 41} (1978), 601.
\bibitem{burz} J. Burzlaff, Phys. Rev. {\bf D 23} (1981), 1329.
\bibitem{kunz} J. Kunz and D. Masak, Phys. Lett. {\bf B 196} (1987), 513. 
\bibitem{IS}
T. Ioannidou and P.M. Sutcliffe, Phys. Rev. {\bf D 60} (1999), 105009.
\bibitem{br} F. A. Bais and R. J.  Russell, 
Phys. Rev. {\bf D 11} (1975), 2692; Erratum-ibid. {\bf D 12} (1975), 3368.
\bibitem{bfm} P. Breitenlohner, P. Forgacs and D.  Maison, Nucl. Phys. 
{\bf B 383} (1992), 357;
{\bf B 442} (1995), 126. 
\bibitem{bp} Y. Brihaye and B.M.A.G. Piette, Phys. Rev.  {\bf D 64} (2001), 084010. 
\bibitem{bh} Y. Brihaye and B. Hartmann, Phys. Rev.  {\bf D 67} (2003), 044001.
\bibitem{manton} C. J. Houghton, N. S. Manton and P. M. Sutcliffe, Nucl. Phys. 
{\bf B 510} (1999), 507.
\bibitem{Za}
W. J. Zakrzewski, {\it Low dimensional sigma models} (IOP, 1989). 
\bibitem{piette}T. Ioannidou, B. Piette and W. J. Zakrzewski, 
J. Math. Phys. {\bf 40 } (1999), 6353.
\bibitem{acr} U. Ascher, J. Christiansen and R. D. Russell, Math. Comput. {\bf 33},
(1979), 659; ACM Trans. Math. Softw. {\bf 7} (1981), 209.
\bibitem{bhiz} Y. Brihaye, B. Hartmann, T. Ioannidou and W. J. Zakrzewski,
in preparation.
\bibitem{lw} A. Lue and E. J. Weinberg, Phys. Rev. {\bf D60} (1999) 084025;
Y. Brihaye, B. Hartmann and J. Kunz, Phys. Rev. {\bf D 62} (2000), 044008.
\end{thebibliography}
\end{document}